\documentclass[12pt]{article}

\usepackage{epsfig}

\begin{document}

\begin{center}
{\Large \bf Effects of curvature and interactions on the  
dynamics of the deconfinement phase transition.}

\vskip .5cm
{\bf Deepak Chandra \footnote{ S. G. T. B. Khalsa College,
University of Delhi, Delhi-110007, India.} and Ashok Goyal
 \footnote{E--mail:agoyal@ducos.ernet.in }}
\\
{\em  Department of Physics and Astrophysics,}\\
{\em  University of Delhi, Delhi-110007, India.}\\
{\em  and Inter University Center for Astronomy and Astrophysics,}\\
{\em  Ganeshkhind, Pune-411007, India.}\\

\end{center}
\begin{abstract}
\noindent  We study the dynamics of first-order cofinement-deconfinement
phase transition through nucleation of hadronic bubbles in an expanding
 quark gluon plasma in the context of heavy ion collisions for 
interacting quark and hadron gas and by incorporating the effects of 
curvature energy. We find that the interactions reduce the delay in the
phase transition whereas the curvature energy has a mixed behavior. In contrast
to the case of early Universe phase transition, here lower values of
 surface tension increase the supercooling and slow down the hadronization process.
 Higher values of bag pressure 
tend to speed up the transition. Another interesting feature is the start 
of the hadronization process as soon as the QGP is created.  

\end{abstract}
\begin{section}{Introduction}
\par
 The ongoing and planned experiments involving heavy ion collisions 
at RHIC, CERN's SPS and at LHC are aimed at studying the behavior of quantum
 chromodynamics (QCD) at high energies. It is believed that at these high 
energies {\cite{bjo}} (100 GeV/nucleon at RHIC and 3 TeV/nucleon at LHC in the c.m. frame)
 a hot baryon free plasma of quarks and gluons (QGP) is expected to be
 created even as at lesser energies, nuclear collisions at Brookhaven's AGS
 and at SPS show large stopping of the nuclei and hint at the baryon rich
 matter. The QGP plasma will soon undergo a phase transition to hadronic 
matter. The phase transition from QGP to hadron resonance gas (HRG) has been
 extensively studied in the context of the early Universe where the time scale
 is of the order of $10^{-6}-10^{-7}$ s [2-4]. Perturbative finite temperature QCD
 methods along with relativistic transport theory have been used to study the
 early stages of the nuclear collisions by computer simulations [5,6]. The initial
 quantitative picture that emerges is that of a QGP fireball of
 $\sim 150 fm^3$ created at the initial temperature of $\sim 300-350$ MeV,
 which is about twice the expected critical temperature $T_c$. Studies of 
quark-hadron phase transition in the early Universe, in heavy ion collisions
 and in high density nuclear matter in the core of compact stars {\cite{kap}} point to the
 importance of interactions in the two phases. It has been shown that unless
 hadronic interactions are taken into account, the hadronic phase again
 becomes thermodynamically the preferred phase at high temperature and that
 QCD interactions are important in the QGP phase {\cite{sing}} making the critical parameters
 depend on the interactions. The lattice results also point to the departure
 of the QGP equation of state (EOS) from the ideal gas EOS even at 
temperatures well above $T_c$. However there still remain uncertainties in 
the equation of state in both the confined and the deconfined phases. It has
 also been pointed out that in situations in which nucleation takes place, the
 curvature energy $8\pi\gamma r$, a term in the free energy in addition to the
 surface energy term $4\pi r^2\sigma$ (where $\sigma$ and $\gamma$ are the
 surface and curvature energy densities respectively) also plays an important
 role and should be kept in the calculation of the nucleation rate [9-12]. It is in 
this context that the above experiments will go a long way in establishing the
 dynamics of the phase transition and the final picture that will emerge would
 be of great interest in astrophysics and cosmology.
If the phase transition from quark gluon plasma to confined hadronic matter
is a first order thermodynamic phase transition, the transition would proceed
 through the nucleation of hadronic bubbles. Conventionally, the QGP would 
supercool to
 temperature below $T_c$ only then the nucleation starts. The bubbles greater
 than the critical size would expand driving the QGP phase into smaller and
 smaller regions thereby converting more and more matter into the hadronic 
phase. The expansion of the bubble would be accompanied by the release of
 latent heat which will reheat the plasma shutting off further nucleation.
 If the phase transition has to be completed, the latent heat must be gotten
 rid of. Such studies have been made in the context of the early Universe 
where it has been shown that phase transition proceeds in the expanding 
Universe after the Universe supercools to a temperature below $T_c$, the
 expanding bubbles then reheat the Universe expelling the QGP phase till the
 entire Universe is converted to the hadronic phase. The time of completion
 and the degree of supercooling depends on the EOS and the parameters like the
 Bag pressure and the surface tension {\cite{cha}}.
\par
 Csernai and Kapusta {\cite{csernai}} have applied a
 recently computed  nucleation rate to a first order phase transition in a 
set of rate equations to study the time evolution of QGP in heavy ion 
collisions. Based on Bjorken hydrodynamics and on current parameter values, 
they find the transition generates about 30 percent extra entropy and also a 
time delay of $11fm/c$ in completion of the transition. They also estimate that
 the system should supercool about 20 percent below $T_c$ before nucleation 
of hadronic bubbles is rapid enough to start reheating the system.
Csernai and Kapusta for the purposes of illustration used noninteacting system
 of hadrons and QGP. They modelled the hadronic phase by a massless gas of
 pions and the plasma phase by a gas of gluons and massless quarks of two 
flavors. In the light of the importance of taking interactions both in the
 hadronic as well as in the plasma phase for purposes of investigating phase
 transitions at high energies, temperature and densities as discussed above,
 we calculate the rate of formation of the hadron bubbles for interacting QGP
 and hadron resonance gas. It is our endeavour here to examine closely the
 effects of interactions on the dynamics of phase transition as the hot 
plasma of quarks and gluons is created and the temperature drops to $T_c$ 
where a phase mixture of QGP and HRG develops. For this purpose QGP is treated
 as a gas of massless $u$,$d$ quarks, massive $s$ quarks and massless gluons. QCD
 interactions are treated perturbatively to order $g^3$ and the long range
 confinement effects are parameterized by the bag pressure $B$. For the hadron
 gas we use the known spectrum of low lying baryons and meson resonances and
 the repulsive interactions between them by incorporating the hard core 
excluded volume effects. In the literature hadronic interactions have been taken
 in many different ways (see Ref.7 for example), hadronic interactions can be accounted 
for through an average mean field repulsive potential arising from the Reid 
potential and $\pi - \pi$ effective interactions or alternatively, repulsive 
effects can also be included by modifying the chemical potential as done by 
Kapusta et al. in a simple method of implementing the mean field theory. One 
could also use the full fledged framework of relativistic mean field theory
 itself. It has however been shown by Goyal et al.{\cite{kap}} that in the baryon free 
case , all the interaction schemes mentioned above give roughly the same 
value of the hadronic pressure, there are however, marked deviations at very 
high $T$ and/or at high $\mu_B$ regime.
\par
The effect of curvature energy term on the nucleation of hadronic bubbles 
has been studied earlier {\cite{mard}}. It has been found that it leads to a minima in 
the free energy in addition to a maxima below the critical temperature 
and also a minima above $T_c$, 
giving rise to spontaneous nucleation of equilibrium sized bubbles. This may 
contribute to the completion of the phase transition and has interesting 
consequences. 
We study here the effect of curvature on the dynamics of the 
phase transition for several values of the parameters and discuss their 
consequences.
\par
The plan of the paper is as follows. In section 2 we discuss the 
bubble nucleation along with the importance of the prefactor which plays a 
crucial role in guiding the behavior of the nucleation rate. It turns out 
that the exponential factor is of less relevance here unlike the case of 
early Universe where exponent dominates the behavior of the nucleation rate.
 We also study here the effect of curvature on the nucleation rate and give
 the two rate equations to study the time evolution of the transition for 
different parameter values. In section 3 we discuss our result and compare them
 with the case when interactions and the curvature energy are not 
present. We also look at the supercooling that takes place and evaluate 
the behavior of the system with different parameter values.
 Finally in section 4 we give our conclusions.
\end{section}
\begin{section}{Bubble nucleation}
\indent When the QGP cools through the critical temperature $T_c$, the new hadron 
phase becomes preferred. Energetically the new phase remains unfavourable as
 there is surface free energy associated with the QGP and the hadron gas 
interface. Creation of bubbles of the new phase are thus unfavourable and all
 nucleated bubbles with radius less than the critical radius die out but those
 with radii grater than the critical radius expand untill they coalesce with
 each other. Thus supercooling occurs before the new hadron phase actually
 appears and takes over. Consequently reheating takes place due to the release
 of latent heat. The critical radius is obtained by extremizing the 
thermodynamic work expanded to create a bubble, i.e.
\begin{equation}
W \equiv -\frac{4\pi}{3}r^3(P_h - P_q) + 4\pi r^2\sigma - 8\pi(\gamma_q - \gamma_h)
r,
\end{equation}
where $P_h$ and $P_q$ are the pressures in the hadronic and quark phase 
respectively, $\sigma$ is the surface tension and $\gamma = (\gamma_q -
 \gamma_h)$ is the curvature coefficient and has been estimated in {\cite{mard}}.
Following Mardor and Svetitsky the curvature energy in the MIT model for massless 
quarks is given by
\begin{equation}
E_c=\frac {gr} {3\pi } \int_{0}^{\infty } dkk \{ 1+exp( k-\mu_B )/T \}^{-1}
\end{equation}
where $g$ is the statistical weight and $r$ the radius. We get
$\gamma \simeq \gamma_q \equiv \frac {T^2} {16}$, where we have assumed
$\gamma _h << \gamma _q$ following {\cite{hor}}  and $\mu _B$ to be negligible 
compared to $T$. According to Madsen {\cite{mads}} , the assumption of zero strange-quark
 mass in the curvature term should be relaxed but presently no calculation
of the curvature term for massive quarks exists. The surface tension is mainly
due to the finite strange quark mass as the massless particles do not contribute
to the surface term in the free energy.
 The critical radius is obtained by putting $\frac{\partial W}{\partial r} 
= 0$ and we get an extremum below $T_c$ given as
\begin{equation}
r_{c+} = \frac{\sigma}{\Delta P}(1+\sqrt{1-\beta})
\end{equation}
where $\Delta P = P_h-P_q$ and $\beta = \frac{2\Delta P}{\sigma^2} {\gamma }$.
 Since $\beta$ is positive definite, a real solution exists only if
 $\beta \leq 1$. This critical radius is actually a maximum. There is 
another extremum solution (a minima) given by the critical radius
\begin{equation}
r_{c-} = \frac{\sigma}{\Delta P}(1-\sqrt{1-\beta})
\end{equation}
These bubbles correspond to equilibrium bubbles which stabilse to their critical
size $r_{c-}$ and do not grow.
The free energies of the critical sized bubbles are given by
\begin{equation}
W_{c+} = \frac{4\pi \sigma^3}{3\Delta P^2}[2+2(1-\beta)^{3/2}-3\beta]
\end{equation}
and
\begin{equation}
W_{c-} = \frac{4\pi \sigma^3}{3\Delta P^2}[2-2(1-\beta)^{3/2}-3\beta]
\end{equation}
respectively for the two critical radii.
$\beta$ decides the effect of the curvature term and decreases the critical 
radius for the expanding bubbles.
The existence of a critical radius corresponding to the minimum of 
the free energy is a consequence of the curvature term.
In fact there is a extremum in the free energy even above the critical temperature
for all starting temperatures of the plasma. This extremum is also a minimum 
and the corresponding critical radius and the free energy are given by
\begin{equation}
r_{c+>} = \frac{\sigma}{\Delta P}(-1+\sqrt{1+\beta})
\end{equation}
\begin{equation}
W_{c+>} = \frac{4\pi \sigma^3}{3\Delta P^2}[2-2(1+\beta)^{3/2}+3\beta]
\end{equation}
Significantly, the restriction of $\beta \leq 1$ is not present for these bubbles.
This feature of the curvature term ensures that the phase transition actually
begins well above the critical temperature (as soon as the plasma is formed) 
by the nucleation of equilibrium 
sized bubbles of the hadron phase, and this continues till below the critical 
temperature when the expanding bubbles also join them and start the rapid 
nucleation and expansion of the hadron phase. As pointed out earlier, below $T_c$ 
if  $\beta$ becomes greater than one, nucleation of both types of critical 
bubbles dissapear. The phase transition can now take place by the expansion of all
the already nucleated critical bubbles $r_{c+}$, $r_{c-}$ and $r_{c+>}$. This is
because the critical radius $r_c$ and the critical free energy $W_c$ necessary
for creation of growing bubbles become zero.
This essentially means that the phase transition now gets rolling 
and completes fast. Sooner $\beta$ becomes greater than $1$, faster will this  
process dominate the phase tranisition. As we will see , for certain reasonable 
values of the parameters ($B$ and $\sigma$) $\beta $ becomes greater than $1$
 almost near $T_c$ causing unrestrained expansion of all the bubbles (ones above
$T_c$ as well as whatever exists below it) and completion of the phase transition.
\par 
The bubble nucleation rate(number of bubbles formed per unit time per unit 
volume) at temperature $T$ is given by
\begin{equation}
I = I_0e^{-W_c/T}
\end{equation}
where $I_0$ is the prefactor having dimensions of $T^4$. The prefactor used 
traditionally in the early Universe studies is given by 
$I_0=(W_c/2\pi T)^{3/2}T^4$. Csernai and Kapusta {\cite{csernai}} have recently calculated 
this in a course grain effective field theoretic approximation to QCD and give
$$I_0 = \frac{16}{3\pi}\bigg(\frac{\sigma}{3T}\bigg)^{3/2}\frac
{\sigma \eta_qr_c}{\xi_q^4(\Delta w)^2}$$
where $\eta_q=14.4T^3$ is the shear viscosity in the plasma phase, $\xi_q$ 
is a correlation length of order 0.7 fm in the plasma phase, and $\Delta w$ 
is the difference in the enthalpy densities of the two phases. This 
nucleation rate is limited by the ability of dissipative process to carry
 latent heat away from the bubbles surface as indicated by the dependence on
 the viscosity. The above expression however gets modifed in the present case 
because of the presence of the equilibrium bubbles. We now have different $I$'s
depending on which kind of bubbles are being discussed, namely $I_+$, $I_-$ and
$I_{+>}$ corresponding to critical radiuses $r_{c+}$, $r_{c-}$ and $r_{c+>}$ 
respectively.
\par 
 The pressure in the QGP phase can be calculated by using the
 thermodynamic potential given in Kapusta and Shuryak {\cite{kap}} . For the hadron phase we
 use the known masses of the low lying 33 baryons and 45 mesons whose masses 
and degeneracy factors are taken from the Particle Data Group Summary {\cite{dat}}. The 
hadronic pressure and number densities for non-interacting point baryons and 
mesons are given by the usual thermodynamic relations. One of the simplest 
ways to account for the short range repulsive forces as discussed above is by
 considering the finite volume of the baryons that modify the space available
 for occupation in a manner akin to that of a Van der Waal's equation. We take
 baryons and antibaryons to have the same size as protons, given by 
$V_p=m_p/4B$ where $B$ is the bag pressure. A radius $r_p$ lying between 0.6 fm
 and 0.8 fm is often used {\cite{kap}} . The hadronic pressure and baryon densities 
corrected for finite volume effects are now given by
\begin{equation}
P_h = \frac{\sum_b P_b^{pt}}{1 + \sum_b n_b^{pt}V_p} + \frac{\sum_{\bar b}
P_{\bar b}^{pt}}{1 + \sum_{\bar b}n^{pt}_{\bar b}V_p} + \sum_mP_m^{pt}
\end{equation}
\begin{equation}
n_B = \frac{\sum_b n_b^{pt}}{1 + \sum_b n_b^{pt}V_p} + \frac{\sum_{\bar b}
n^{pt}_{\bar b}}{1 + \sum_{\bar b}n^{pt}_{\bar b} V_p}
\end{equation}
where $b$, $\bar b$ and m stand for baryons, anti-baryons and mesons 
respectively. The pressure equilibrium between the two phases sets $T_c$ 
which is decided by the parameters in the theory, essentially by the parameters
 $B$ and $\sigma$.
With nucleation rate given by eq.(9), one would like to know the fraction of 
space $h(t)$ which is in the hadronic phase at the given time t as measured in a local
 comoving frame of the expanding system. Guth and Weinberg {\cite{guth}} have proposed a 
formula for $h(t)$ in the cosmological first order phase transition. Csernai 
and Kapusta {\cite{csernai}} have given a more accurate kinetic equation than the one used by
 Guth and Weinberg and it allows for the transition to complete.
If the system cools to $T_c$ at time $t_c$, then at some later time $t$, the 
fraction of space which has been converted to hadronic gas is given by them as
\begin{equation}
h(t) = \int_{t_c}^t dt' I(T(t'))\{1-h(t')\}V(t',t)
\end{equation}
where $V(t,t')$ is the volume of a bubble at time $t$ which had been 
nucleated at the earlier time $t'$; taking bubble growth into account.
 Here we have neglected bubble collisions and fusion. 
The above expression however gets modifed in the present case 
because of the presence of the equilibrium bubbles. Now we have an additional
 contribution to the hadron fraction from the bubbles that are nucleated above
 the critical temperature plus those that are nucleated below the critical
 temperature and do not grow in size (unless $\beta >1$). 
So, the new expression is now given by
$$h(t) = \int_{t_c}^t dt' I_+(T(t'))\{1-h(t')\}V(t',t)$$
$$+\int_{t_c}^t dt' I_-(T(t'))\{1-h(t')\}\frac{4\pi}{3}r_{c-}^3(T(t'))$$
\begin{equation}
+\int_{t_0}^{t_c} dt' I_{+>}(T(t'))\{1-h(t')\}\frac{4\pi}{3}r_{c+>}^3(T(t')) 
\end{equation}
The dynamical equation which couples time evolution of temperature to $h(t)$ is
 obtained by using the longitudinal scaling hydrodynamics of Bjorken. It is 
given by 
\begin{equation}
\frac{de}{dt} = -\frac{w}{t}
\end{equation}
where $e(T)$ and $w(T)$ are the energy and enthalpy densities respectively being 
given by $e(T)=h(t)e_h(t) + (1-h(t))e_q(t)$, $e_h$ and $e_q$ being the hadron
 and quark energy densities at time t respectively, likewise for the enthalpy 
density.
Following {\cite{mill}} the growth velocity of bubbles is taken to be 
$v(T)=v_0(1-T/T_c)^{3/2}$ where $v_0=3c$ for $T>2/3T_c$. This gives 
$v<c/\sqrt3$, the speed of sound for a massless gas. We have assumed that 
if supercooling takes $T$ below $2/3T_c$ then $v$ is given by $c/\sqrt3$. This 
ensures that closer $T$ is to $T_c$ slower the bubbles grow.
 The volume of an expanding bubble at time $t$ is then given by
\begin{equation}
V(t',t) = \frac{4\pi}{3}\{r_{c+}(T(t')) + \int_{t'}^t dt'' 
v(T(t''))\}^3
\end{equation}
In calculations we use the parameters $\xi_q=0.7fm$ and $\eta_q=14.4T^3$ as in {\cite{csernai}}. 
$\sigma$ and $B$ are varied to study the behavior of the phase transition 
as the QGP cools and hadron bubbles are formed.
We solve the two coupled equations (13) and (14) for different values of
the parameters $B$ and $\sigma$.
 The nucleation time is defined as $$\tau ^{-1} = \frac {4\pi } 
{3} r^3_cI$$ neglecting bubble growth. In our case we have a contribution to
 $\tau $ from $I_+$, $I_-$ below $T_c$ and $I_{+>}$ above $T_c$.
\end{section}
\begin{section}{Results and Discussion}
\par
\indent We find that the inclusion of the interactions in both the plasma phase 
and the hadron gas phase lowers the critical temperature, so higher values of the bag 
pressure are required for having reasonable values of $T_c$. For example, 
the value $B^{1/4}=300 MeV$ gives a $T_c \sim 188 MeV$ whereas $B^{1/4}=235 MeV$ gives a 
$T_c \sim 144 MeV$ . We show below plots of free energy vs radius of a hadron 
bubble, nucleation time vs temperature, temperature vs time and hadron fraction 
vs time etc. In fig.1 we plot the free energy of a hadron bubble against it's radius 
 in the QGP plasma. For temperatures just above the critical 
temperature we see that the effect of the curvature term is to produce a minima
in the free energy which is to be contrasted with no extremum without a curvature term.
 Fig.1a pertains to $T=1.01T_c$ and the dashed and the dotted curves are without 
the curvature term. It is clearly seen that equilibrium sized hadron bubbles appear  
even above the critical temperature (solid and long dashed curves). Fig.1b is 
plotted  for a different set of parameters. The lower panel is plotted just below 
the critical 
temperature at $T=0.99T_c$. Here we find that below the 
critical temperature
 the curvature effect produces a minima as well as a maxima in the free energy. 
The maxima bubbles 
grow whereas the minima bubbles reach an equilibrium size. The dashed and the 
dotted curves are without the curvature effect showing the presence of only
 the maximum sized bubbles. Figures 1c and 1d plot the same curves but for a 
different set of the parameters. It is interesting to note that for 
$B^{1/4}=300 MeV$, $\sigma \sim
7 MeV/fm^2$ ($\sigma^{1/3}=65 MeV$) and $T=0.99T_c$ the free energy 
has no extremums, only a
maximum value at zero (long dashed), indicating that critical radius is zero and 
all existing bubbles can now expand to reduce the free energy. In such cases 
$\beta $ becomes greater than 1 soon after $T_c$ is crossed and there 
cannot be nucleation of any fresh bubbles. The expansion of all the existing bubbles 
 ensures completion of the transition.
In Fig.2 we have shown the plot of $T/T_c$ agains time. 
The temperature of the 
QGP plasma decreases with time till $T_c$, when the phase transition conventionally
 begins and supercooling starts. The degree of supercooling  depends on the 
parameters and can be a significant fraction of $T_c$. The important feature
 to see here is that the prefactor $I_0$ is the deciding factor for 
supercooling and the duration of the transition rather than the exponent. 
On the other hand in the early Universe studies, it is the exponent which 
decides the behavior of the transition. In fig.2a the dashed curve 
(without curvature) and the 
solid curve (with curvature) are 
for parameters $B^{1/4}=235 MeV$ and $\sigma =
50 MeV/fm^2$ ($\sigma^{1/3} \sim 125 MeV$). As can be seen 
they overlap, indicating that in this case the curvature term does not
make any difference. This value of the surface tension is rather large and in 
fact exceeds the upper limit set in {\cite{berg}}. However, for the parameters 
$B^{1/4}=235 MeV$ and  
$\sigma =20 MeV/fm^2$ ($\sigma^{1/3} \sim 92 MeV$) there is a delay in the 
completion of the phase transition
 when the $\gamma $ term is included (long dashed) compared to the no 
$\gamma $ (dotted) case. The delay is of the order of 13 fm and is because of $\beta $
 becoming greater than one a little below $T_c$ with consequent stoppage of
 all bubble nucleation. When the $\gamma $ term is not taken, there is no 
such stoppage and continous nucleation ensures faster completion. Figure 2b shows the 
same curves for the parameters $B^{1/4}=235 MeV$, 
$\sigma \sim 7 MeV/fm^2$ and $B^{1/4}=300 MeV$, 
$\sigma \sim 7 MeV/fm^2$. In the latter case, the curvature term 
completly inhibits bubble nucleation below $T_c$ and only the equilibrium 
bubbles formed above $T_c$  expand to complete the phase transition (long dashed).
This corresponds to the case of no extremum in free energy (fig.2b). 
The inclusion of $\gamma $ term in this case actually reduces the time to complete
 the phase transition. This can be contrasted with the case corresponding to 
the other set of parameters in fig.1d and the cases shown in fig.2a. Comparing 
with the no curvature case,
It is seen that 
increasing $B$ speeds up the phase transition whearas decreasing it slows it 
down if the $\gamma $ term is included.
 We also see that the decrease in the surface tension increases the supercooling 
of the plasma by significant amount (can be close to 50 percent). However, increasing 
$B$ reduces supercooling when $\gamma $ is included.
\par
Next we look at the nucleation rate vs $T/T_c$. One can see from these graphs 
(figures 3a and 3b)
 that nucleation of hadronic bubbles can start even above the critical temperature 
when we include the curvature energy. The effect of decreasing $\sigma $ is to
suppress nucleation. We may note that in the early Universe reverse is the 
case, and it happens because there the exponent drives the nucleation rate whearas 
in QGP the prefactor drives the nucleation. The point at which $\beta $ becomes
greater than one is clearly visible when curvature term is included. In figure 3a 
 the long dashed curve terminates at the point where further nucleation stops. It
 is also clear that nucleation starts the moment the plasma is created well above 
the critical temperature if the curvature term is included. Figure 3b shows that 
for smaller surface tension the $\beta $ term becomes greater than one soon after 
crossing through $T_c$ (solid and the long dashed curves). The other two curves 
shown are without the curvature terms and so begin below the critical temperature. 
\par
In figure 4 upper panel the 
critical radius of a bubble is plotted as a function of $T/T_c$ for those bubbles 
which are nucleated at the minima in the free energy. These bubbles in general 
are nucleated above $T_c (r_{c+>})$ as well as below $T_c (r_{c-})$. In fig.4a we see that for the 
parameters $B^{1/4}=235 MeV$, $\sigma =20 MeV/fm^2$ 
the bubbles keep nucleating 
from above the critical temperature to well below it. However for another set 
($B^{1/4}=300 MeV$, $\sigma \sim 7 MeV/fm^2$) these bubbles are no longer nucleated 
 soon after crossing $T_c$. In figure 4b we find this feature in both the 
parameter sets. In all such situations (where $\beta $ becomes greater than 1) 
the completion of the phase transition is ensured by expansion of already 
nucleated bubbles.
The lower panel plots the critical bubble radius $r_{c+}$ corresponding to the maxima 
in the free energy as a function of $T/T_c$ for 
temperatures below the critical temperature ($T<T_c$). We see in both the graphs 
4c and 4d that the critical radius goes down with temperature. In figure 4c we 
see that the critical bubble radius goes to zero when $\beta >1$ (long 
dashed curve having $\gamma \neq 0$). If no curvature term is present then the critical 
size does not go to zero. Even with curvature term, large $\sigma $ does not allow
 critical radius to go to zero (solid curve). Figure 4d plotted for a different set 
of parameters has similar features except that for $B^{1/4}=300 MeV$, 
$\sigma \sim7 MeV/fm^2$ the critical radius below $T_c$ is zero. The only bubbles 
that can expand are the ones already nucleated above $T_c$. 
\par
Finally, figures 5a and 5b show the hadron fraction $h$ as a function of time $t$ 
as the phase transition proceeds to it's completion. Figure 5a shows the overlap
for one parameter set given there (solid and dashed curves)  and clearly shows 
that if we decrease the surface 
tension the phase transition proceeds slowly (long dashed curve) unlike the 
case in early Universe 
studies. The effect of $\gamma $ has a mixed behavior depending on the value of 
the bag pressure. Figure 5b shows that for low bag pressures, the effect 
of $\gamma $ is to delay the phase transition (solid curve) but for 
larger values of $B$ we 
see that the curvature correction actually speeds up the process  of hadronization 
(long dashed curve).
\par
The idealized Maxwell construction with the correction to the QGP as well as the 
the hadron gas that we have studied here shows that the time range of the phase
transition is actually less than what has been given in {\cite{csernai}}. For the same
parameter values as given in {\cite{csernai}} we find that in the ideal case with corrections, 
the phase transition ranges from
2.5 fm to 22 fm whereas without the corrections it ranges from 3 fm to 37 fm.
If we compare our study with this corrected ideal Maxwell construction, we find that 
for $B^{1/4}=235 MeV$ the presence of the curvature term delays the transition to 31 fm for 
$\sigma =50 MeV fm^{-2}$, to 53 fm for $\sigma =20 MeV fm^{-2}$ and to 
more than 60 fm for $\sigma \sim 7 MeV fm^{-2}$ . For $B^{1/4}=300 MeV$ the ideal case streches 
from 1.9 fm to 6.8 fm but curvature introduces a 
delay and the transition extends to 12 fm for 
$\sigma =50 MeV fm^{-2}$, to 16 fm for $\sigma =20 MeV fm^{-2}$ and to 27 fm for 
$\sigma \sim 7 MeV fm^{-2}$. It is interesting to note that the ideal 
Maxwell case range is 1.5 fm to 2.8 fm for $B^{1/4}=350 MeV$, but $\sigma \sim 7 MeV fm^{-2}$ 
with the curvature term produces a range from 1.5 fm to 16 fm.  
\end{section}
\begin{section}{Conclusions}
\par
In conclusion, we have presented in detail the mechanism of the phase transition 
of the QGP fireball from its creation to it's subsequent hadronization in an 
ultrarelativistic nuclear collision. The inclusion of the interactions in both phases 
along with the  
the curvature energy term introduces subtlities that are both interesting and 
quantitative in nature. Broadly, we can summarize that whereas interactions lower 
the critical temperature, the combined effect of interaction and curvature energy 
is to affect both the degree of supercooling 
and the extent of the delay in the phase transition. Smaller values of the surface 
tension can enhance supercooling to almost 50 percent and also increase the delay 
in the hadronization process whereas larger values of the bag pressures speed 
up the hadronization process. $\sigma ^{1/3}=65 MeV$ ($\sigma \sim 7 MeV fm^{-2}$) 
is a reasonable value 
according to the literature and anything above $\sigma ^{1/3}=105 MeV$ 
($\sigma \sim 30 MeV fm^{-2}$) seems to be on the higher side {\cite{berg}}. The critical 
temperature is decided by the bag pressure and the inclusion of interactions in the two 
phases suppresses it. 
$T_c \sim 170-180 MeV$ is favored in the literature, so a bag pressure in excess 
of $300 MeV$ seems to be on the higher side. Inclusion of the curvature effect 
has a mixed behavior in that, for larger values of $B$ the transition is speeded 
up whereas for smaller values of $B$ delay is enhanced. Another interesting 
observation is that the hadronization actually begins the moment the QGP is 
created, through equilibrium sized bubbles and is later joined by the expanding 
bubbles below the critical temperature. For reasonable parameter values the 
nucleation of bubbles below $T_c$ can get stopped when $\beta >1$ and now 
even the equilibrium sized bubbles join the expansion to complete the phase 
transition. Finally, for the ideal Maxwell construction, inclusion of interactions 
reduces the range of the transition but the presence of the curvature term introduces 
a delay in the completion of the transition compared to the ideal case.
\end{section}
\vskip .5cm
\par
\noindent
{\bf Acknowledgements}
\par
\noindent
This work was partially supported under the SERC Scheme of The Department of 
Science and Technology (DST), India. One of us (A.G.) is thankful to U.G.C. 
for partial financial support.
\vskip .5cm
\par
\noindent
{\bf Figure Captions}
\par
\noindent
Fig. 1. Free energy $W$ in units of MeV as a function of bubble radius $r$ in fm.
 In fig.1a Solid (with $\gamma $) and dashed (without $\gamma $) curves are for 
$B^{1/4} =235 MeV$ and $\sigma =
50 MeV fm^{-2}$. Long dashed (with $\gamma $) and dotted (without $\gamma $) curves 
are for $B^{1/4} =235 MeV$ and $\sigma =
20 MeV fm^{-2}$. In fig.1b Solid (with $\gamma $) and dashed (without 
$\gamma $) curves are for $B^{1/4} =235 MeV$ 
and $\sigma \sim 7 MeV fm^{-2}$. Long dashed (with $\gamma $) and dotted (without 
$\gamma $) curves are for $B^{1/4} 
=300 MeV$ and $\sigma \sim 7 MeV fm^{-2}$. Both the curves 1a and 1b are at a temperature 
of $1.01T_c$. In the lower panel the curves 1c and 1d  are labelled as in Fig.1a 
and 1b respectively except 
that both are at a temperature of $.99T_c$.
\vskip .5cm
\par
\noindent 
Fig. 2. Temperature $T/T_c$  as a function of time $t$ in fm. In fig.2a 
and 2b Solid, dashed, dotted
 and long dashed curves are labelled as in fig. 1a and 1b respectively.
\vskip .5cm
\par
\noindent
Fig. 3. Log of the nucleation time $\tau$(Tau) in units of fm/c as a function of 
temperature. The figures 3a and 3b  are labelled as in fig.1a and 1b respectively. 
\vskip .5cm
\par
\noindent
Fig. 4. Upper panel shows critical bubble radius $r_c$ in fm as a function of $T/T_c$. They are the
minimum free energy equilibrium critcal bubbles. The solid and long dashed curves are both 
with $\gamma $ correction. These bubbles do not exist without the curvature 
correction. Figures 4a and 4b are labelled as in fig.1a and 1b respectively.
Lower panel shows critical bubble radius $r_c$ in fm as a function of $T/T_c$. They are the
maximum free energy expanding critcal bubbles. Figures 4c and 4d are labelled as in 
fig.1a and 1b respectively. 
\vskip .5cm
\par 
\noindent
Fig. 5. The hadron fraction $h$ as a function of time $t$ in fm. Curves in fig.5a and 5b 
are labelled as in fig. 1a and 1b respectively.
\vskip 1.cm

\vskip 1.cm
\pagebreak
\begin{figure}[ht]
\vspace*{1cm}
\centerline{
\epsfxsize=5cm\epsfysize=6cm\epsfbox{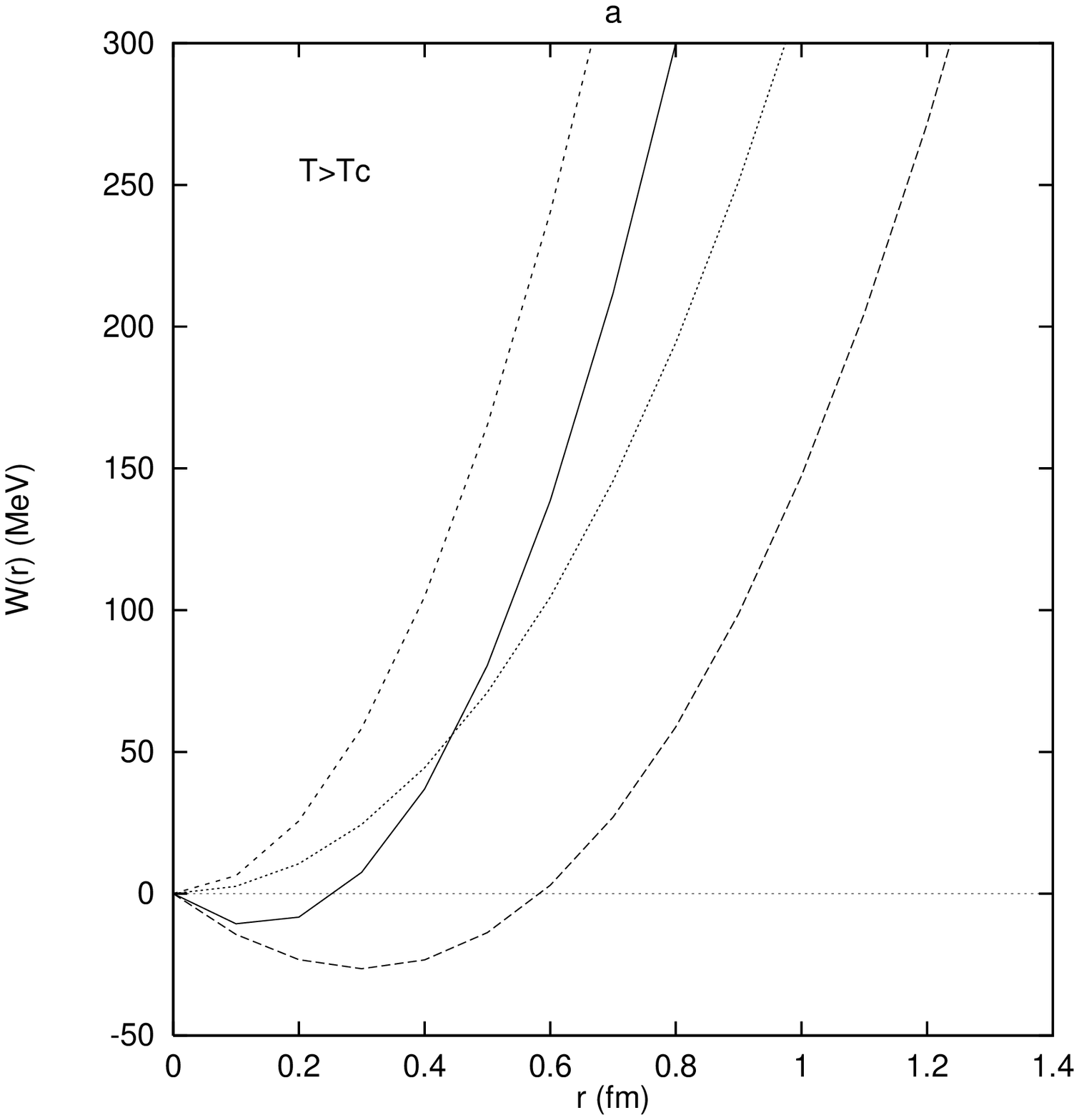}
\hskip 1cm
\epsfxsize=5cm\epsfysize=6cm\epsfbox{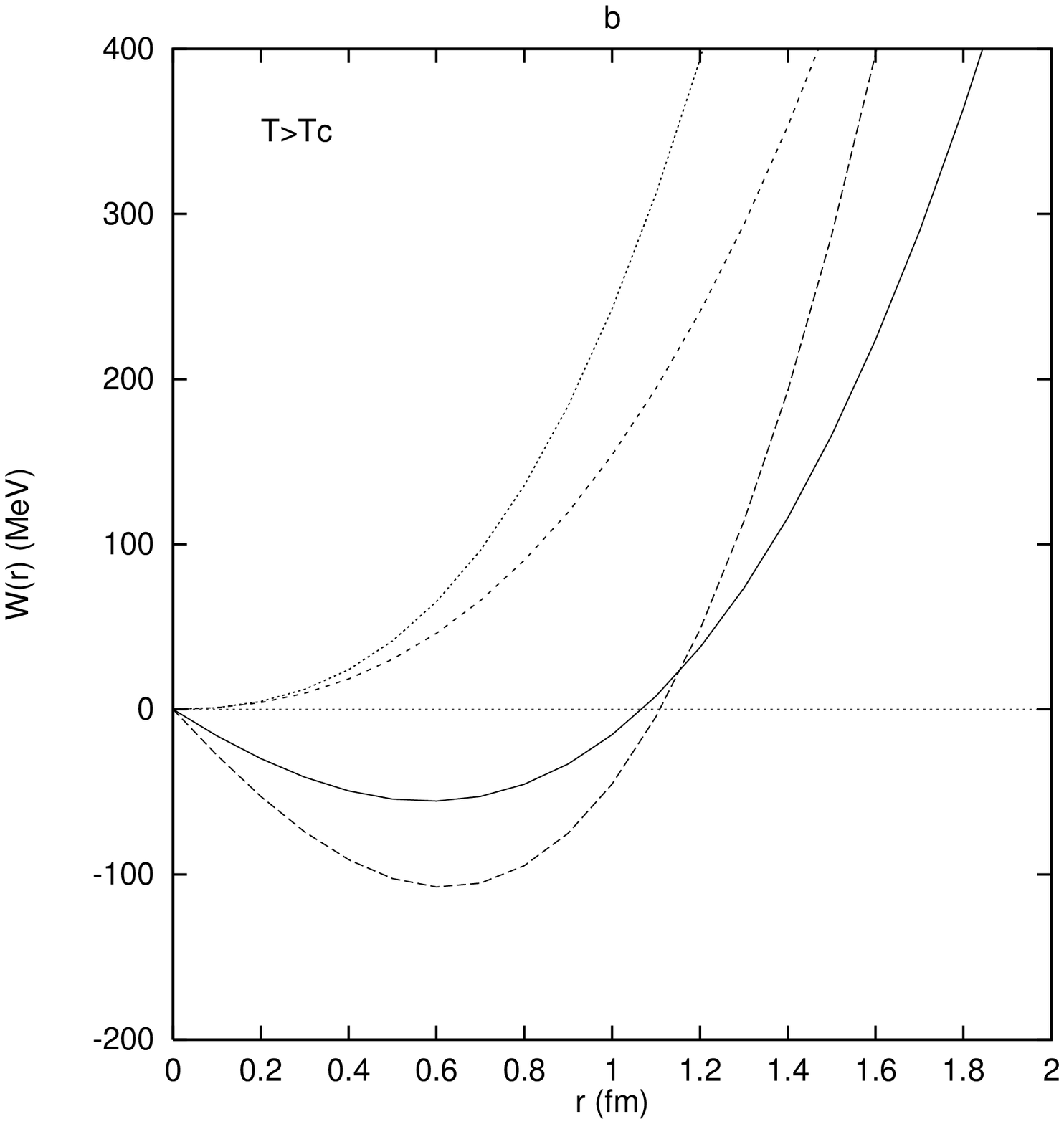}}
\vskip .5cm
\centerline{
\epsfxsize=5cm\epsfysize=6cm\epsfbox{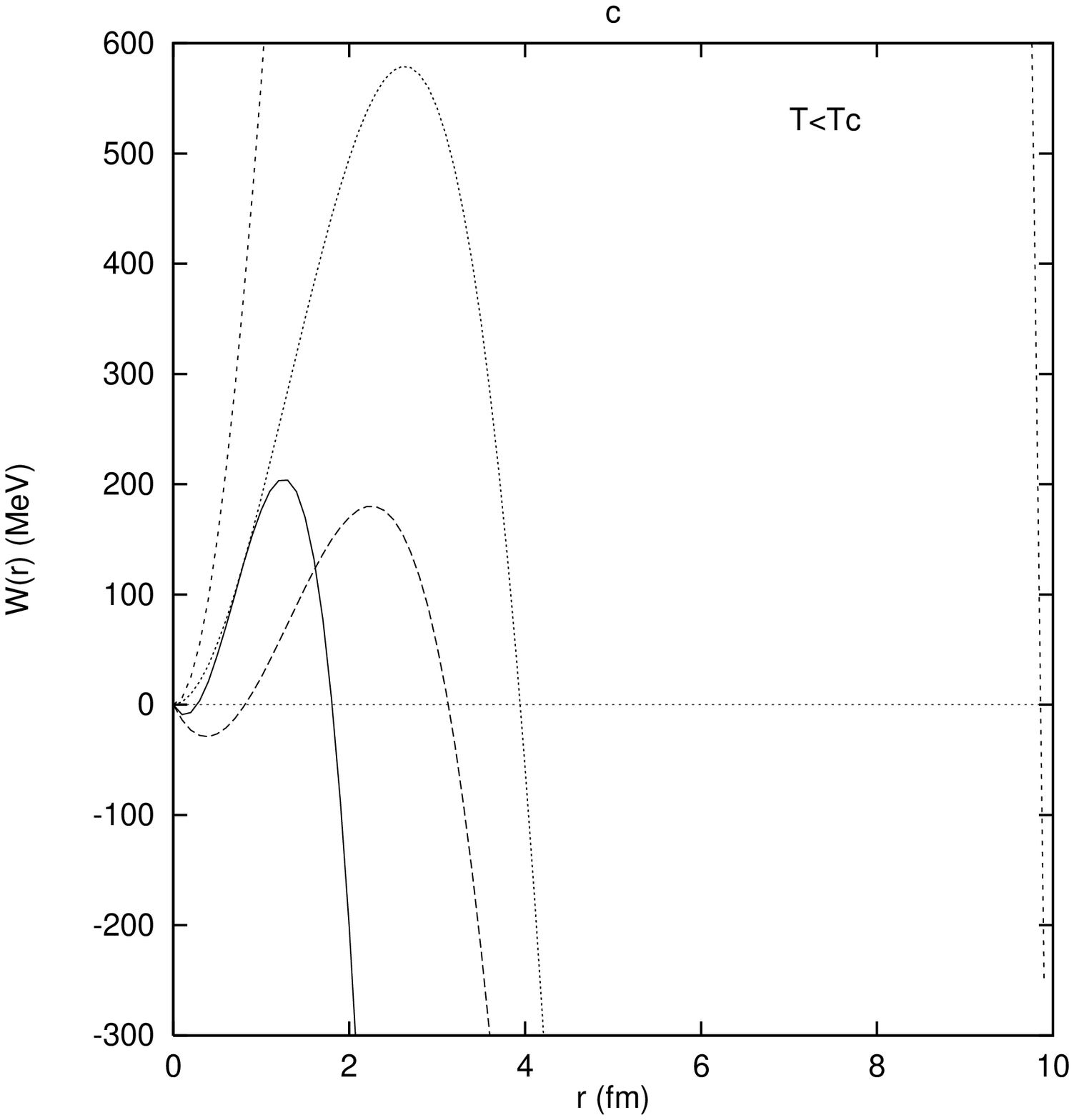}
\hskip 1cm
\epsfxsize=5cm\epsfysize=6cm\epsfbox{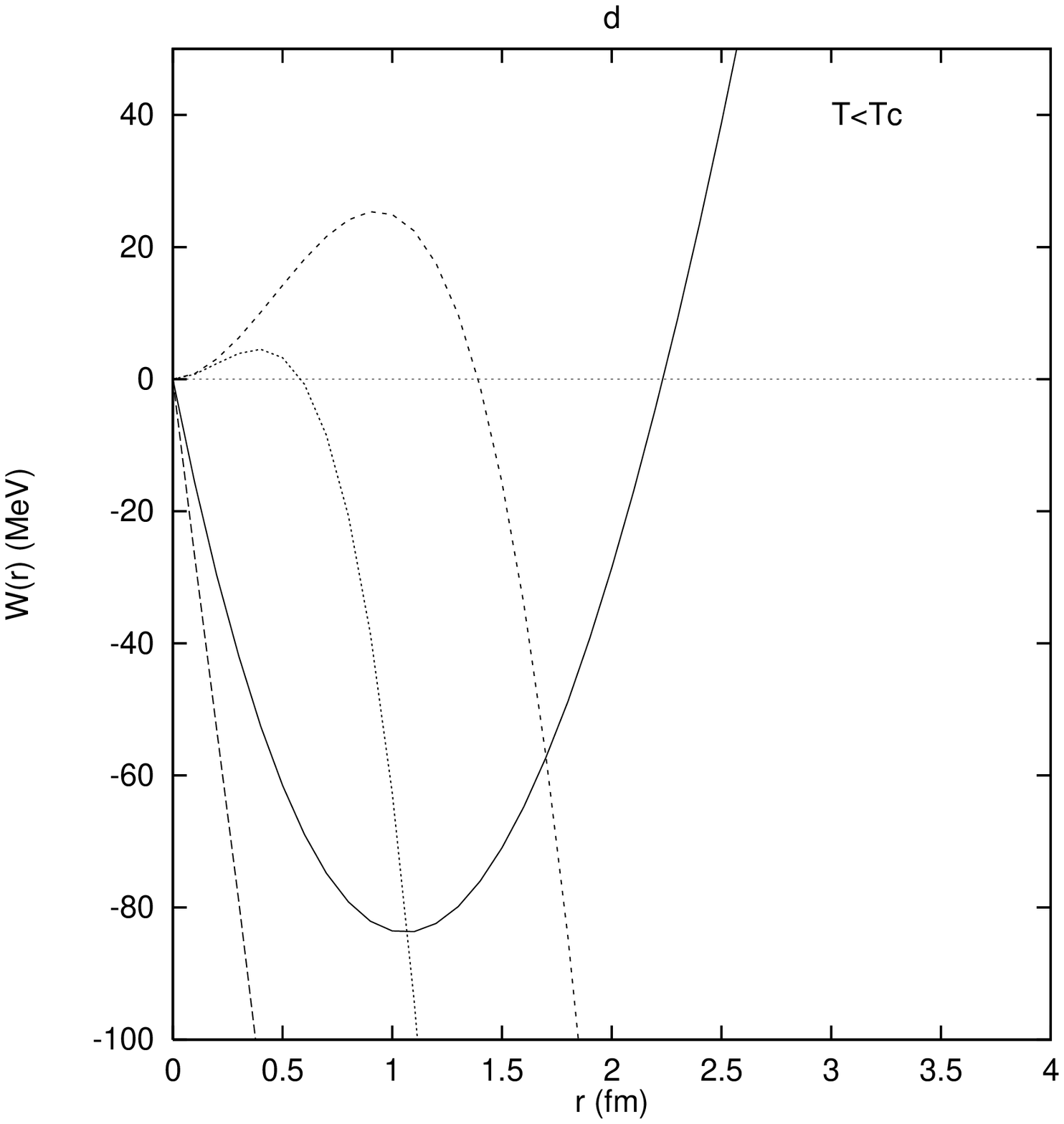}}
\caption{Free energy $W$ in units of MeV as a function of bubble radius $r$ in fm.
 In fig.1a Solid (with $\gamma $) and dashed (without $\gamma $) curves are for 
$B^{1/4} =235 MeV$ and $\sigma =
50 MeV fm^{-2}$. Long dashed (with $\gamma $) and dotted (without $\gamma $) curves 
are for $B^{1/4} =235 MeV$ and $\sigma =
20 MeV fm^{-2}$. In fig.1b Solid (with $\gamma $) and dashed (without 
$\gamma $) curves are for $B^{1/4} =235 MeV$ 
and $\sigma \sim 7 MeV fm^{-2}$. Long dashed (with $\gamma $) and dotted (without 
$\gamma $) curves are for $B^{1/4} 
=300 MeV$ and $\sigma \sim 7 MeV fm^{-2}$. Both the curves 1a and 1b are at a temperature 
of $1.01T_c$. In the lower panel the curves 1c and 1d  are labelled as in Fig.1a 
and 1b respectively except 
that both are at a temperature of $.99T_c$.}
\end{figure}
\pagebreak
\begin{figure}[ht]
\vspace*{1cm}
\centerline{
\epsfxsize=5cm\epsfysize=6cm\epsfbox{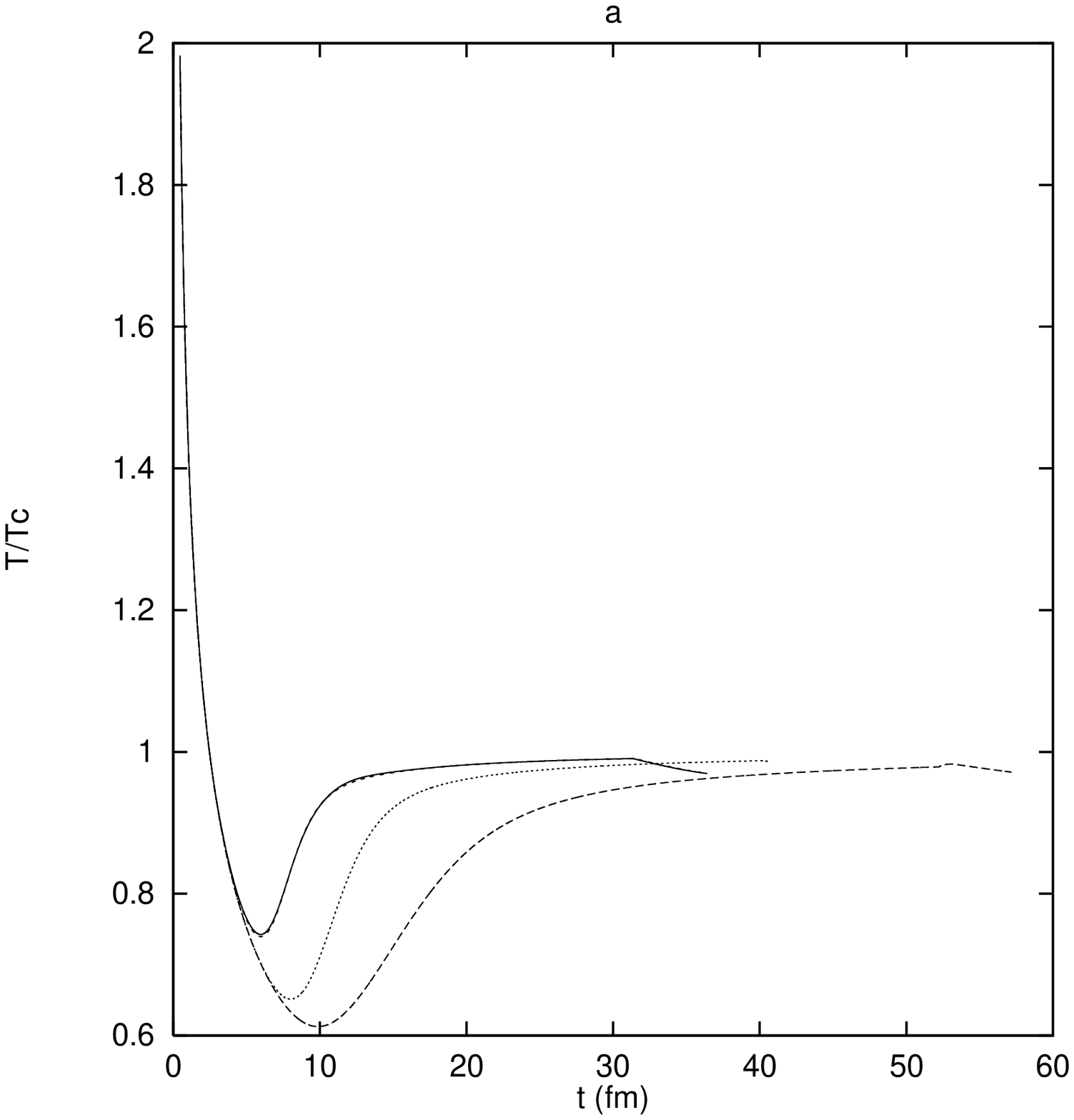}
\hskip 1cm
\epsfxsize=5cm\epsfysize=6cm\epsfbox{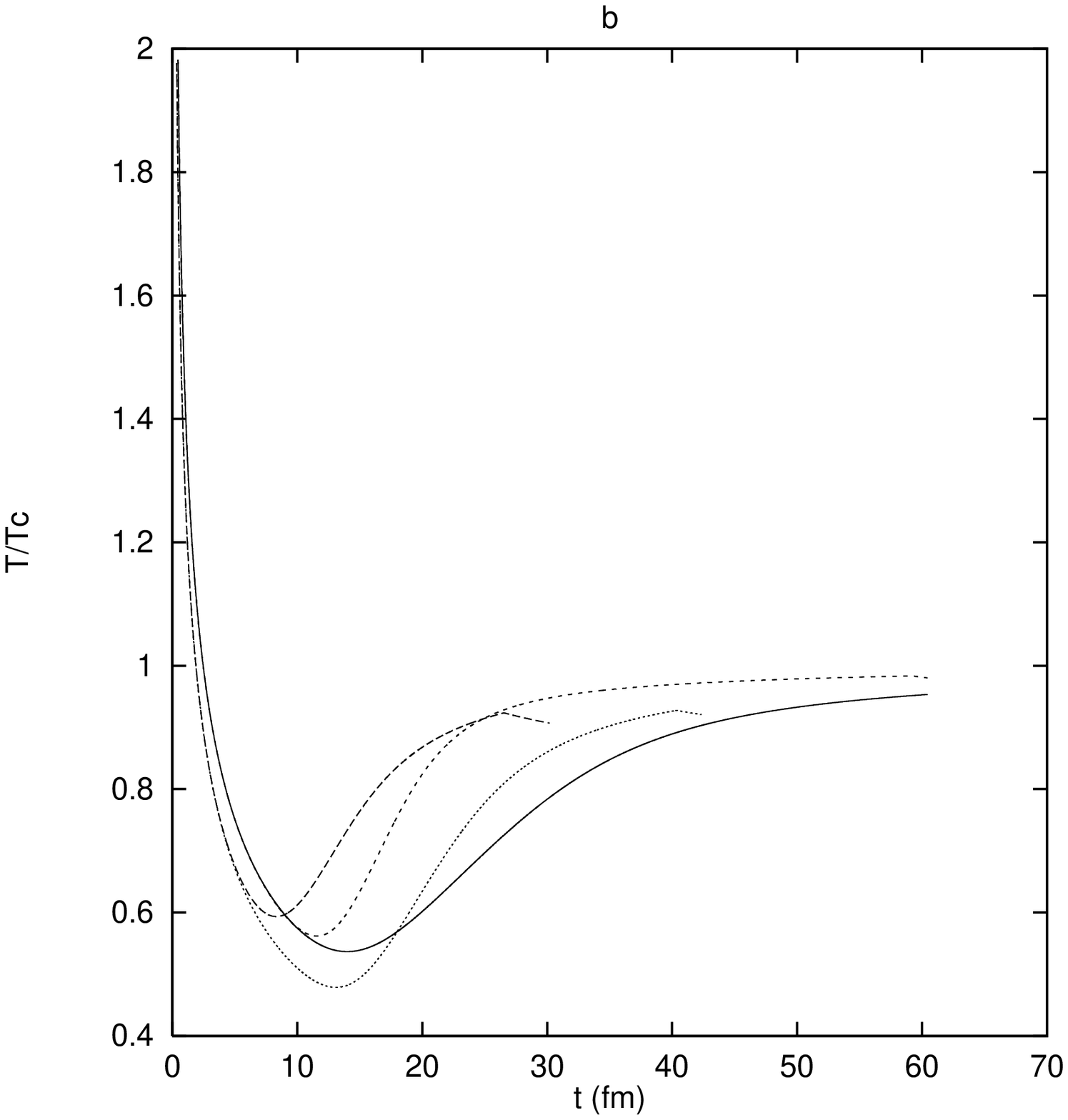}}
\caption{Temperature $T/T_c$  as a function of time $t$ in fm. In fig.2a 
and 2b Solid, dashed, dotted
 and long dashed curves are labelled as in fig. 1a and 1b respectively.}
\end{figure}
\pagebreak
\begin{figure}[ht]
\vspace*{1cm}
\centerline{
\epsfxsize=5cm\epsfysize=6cm\epsfbox{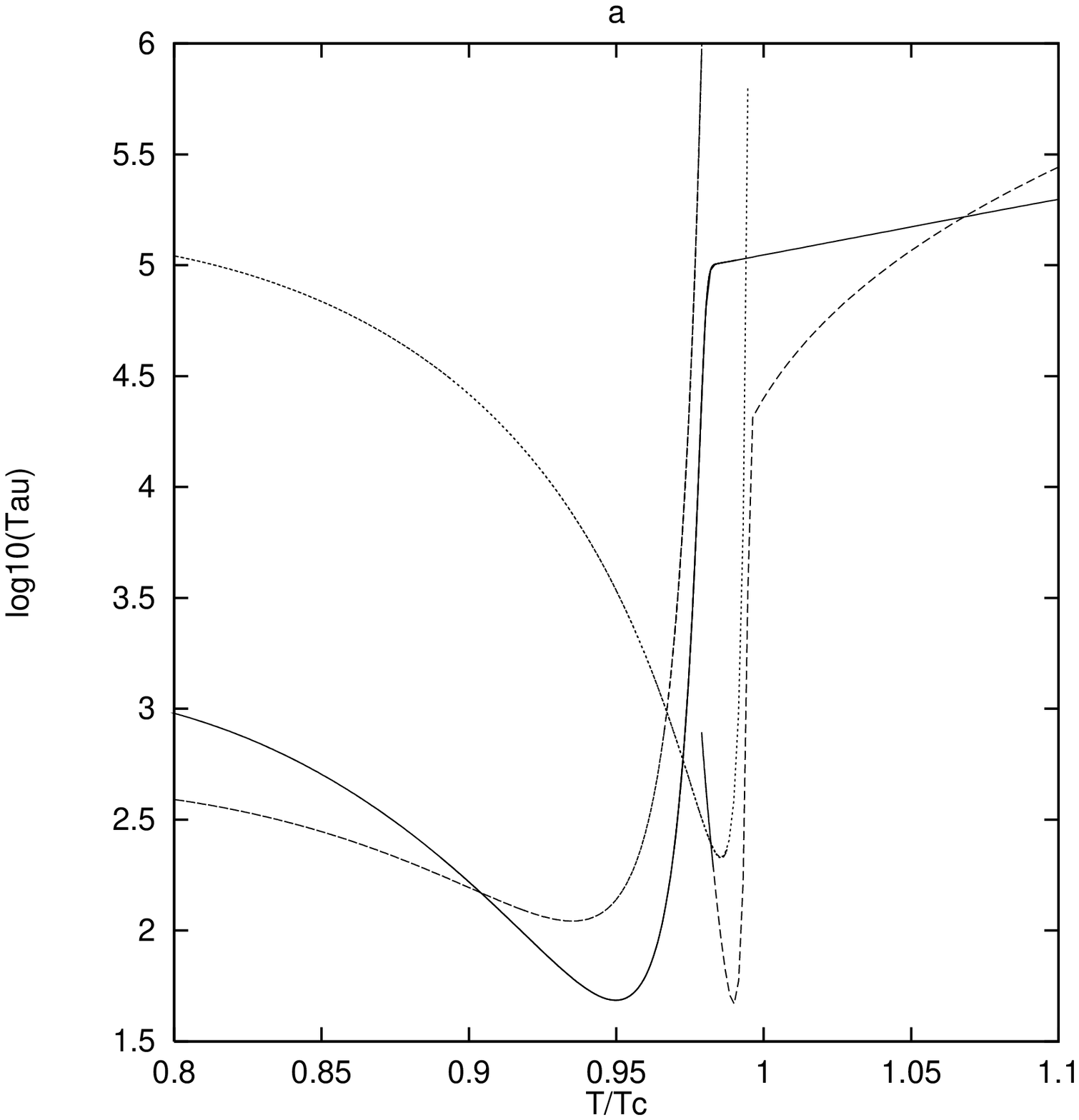}
\hskip 1cm
\epsfxsize=5cm\epsfysize=6cm\epsfbox{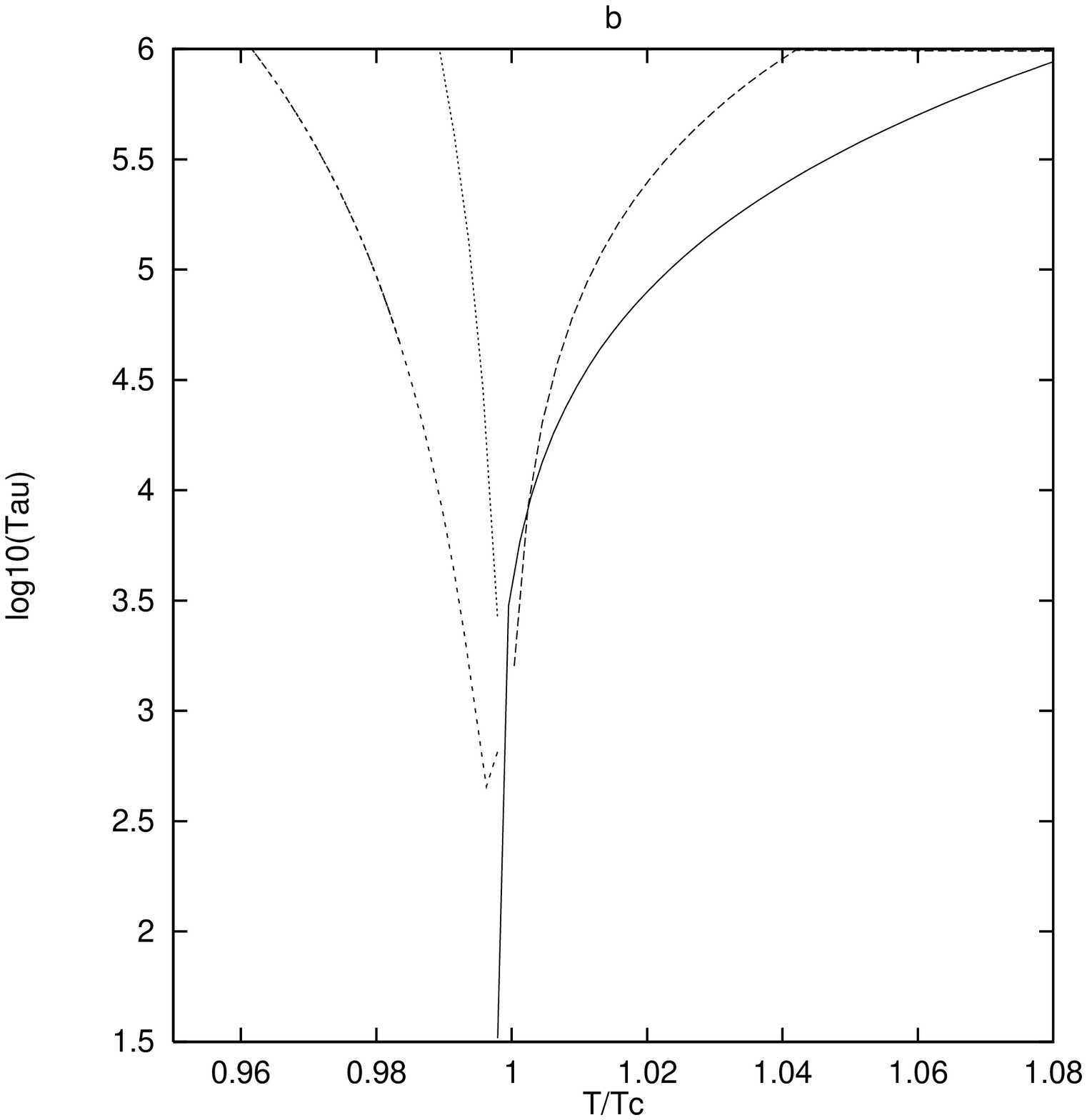}}
\caption{Log of the nucleation time $\tau$(Tau) in units of fm/c as a function of 
temperature. The figures 3a and 3b  are labelled as in fig.1a and 1b respectively.}
\end{figure}
\pagebreak
\begin{figure}[ht]
\vspace*{1cm}
\centerline{
\epsfxsize=5cm\epsfysize=6cm\epsfbox{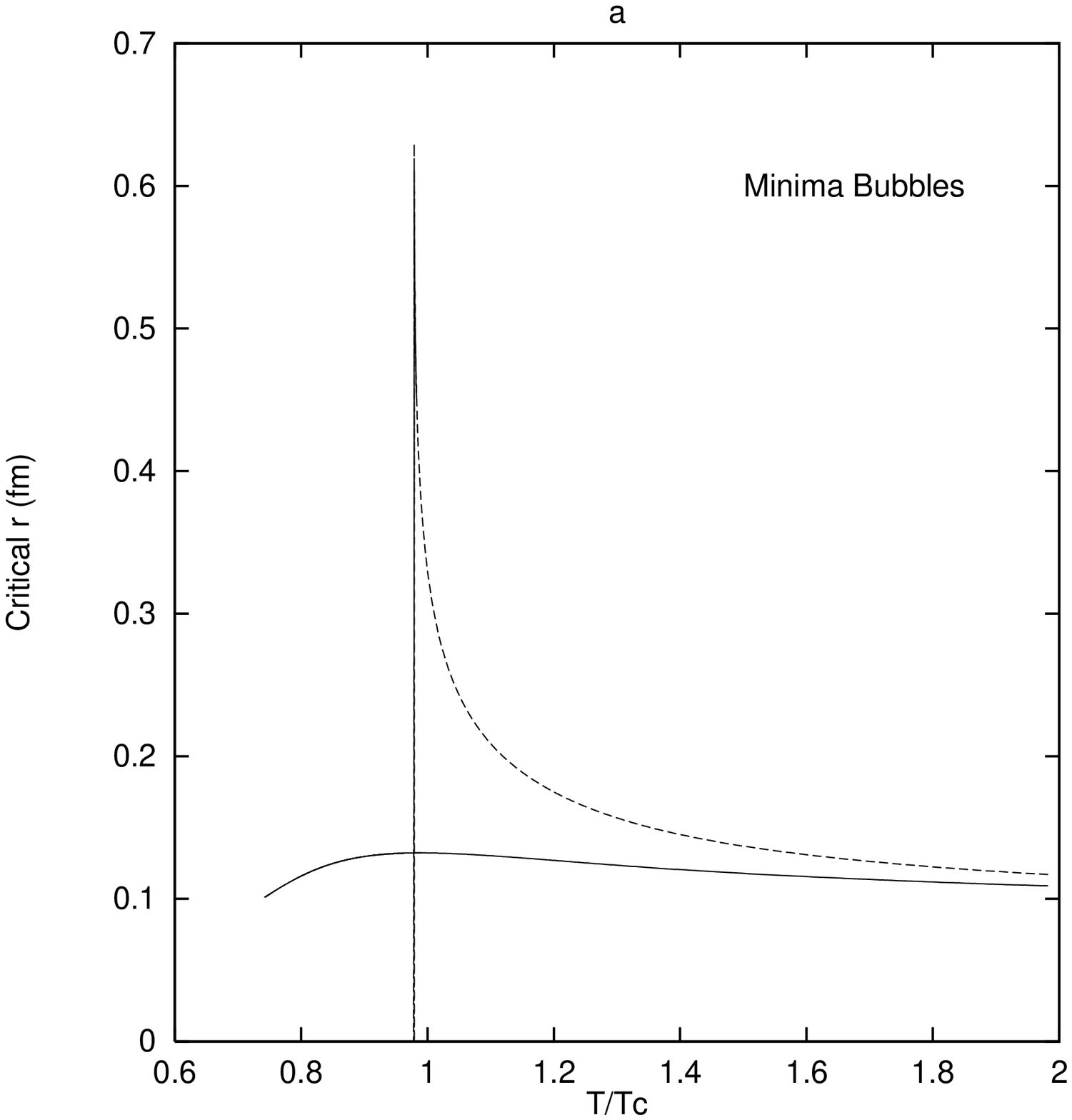}
\hskip 1cm
\epsfxsize=5cm\epsfysize=6cm\epsfbox{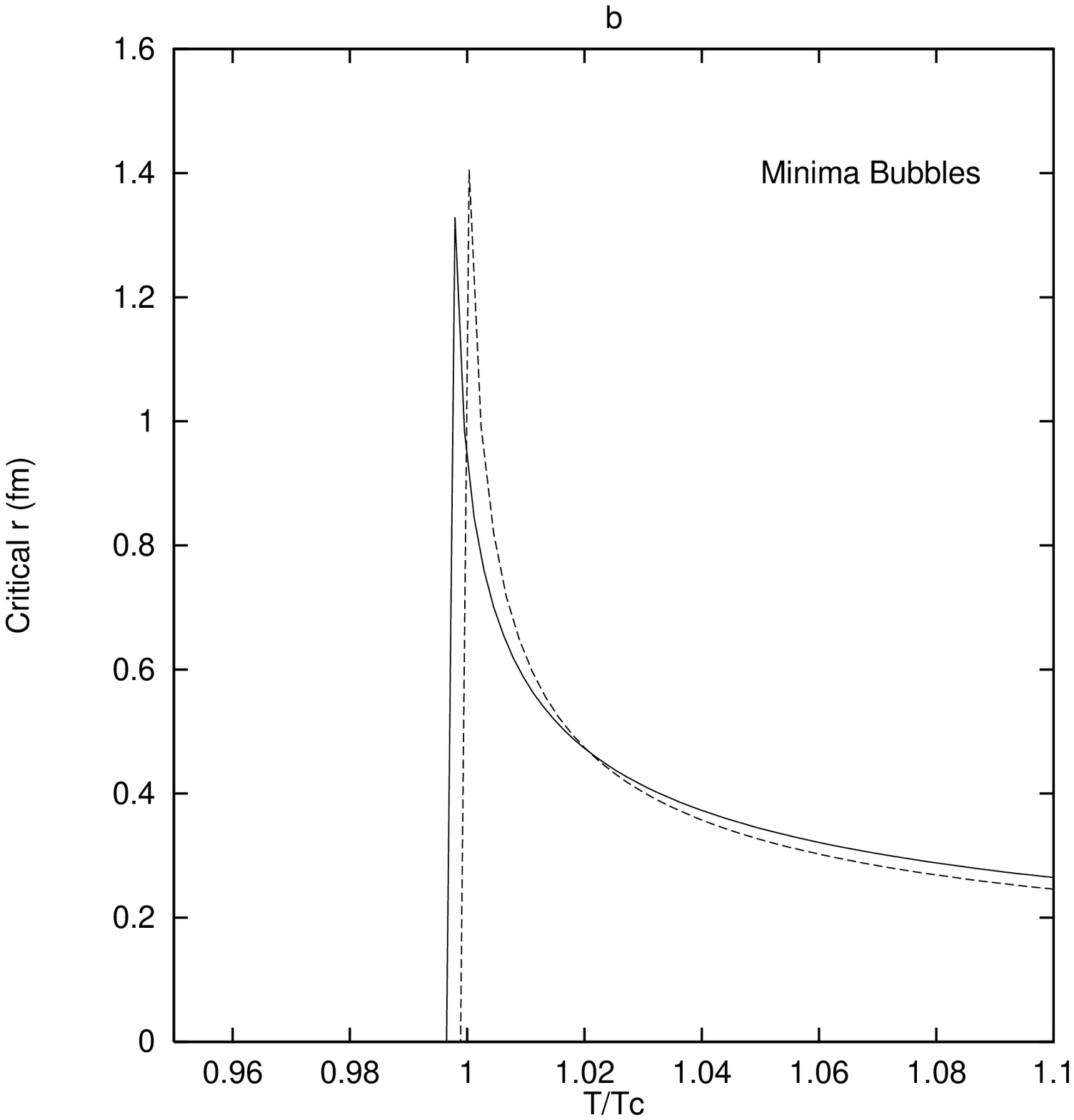}}
\vskip .5cm
\centerline{
\epsfxsize=5cm\epsfysize=6cm\epsfbox{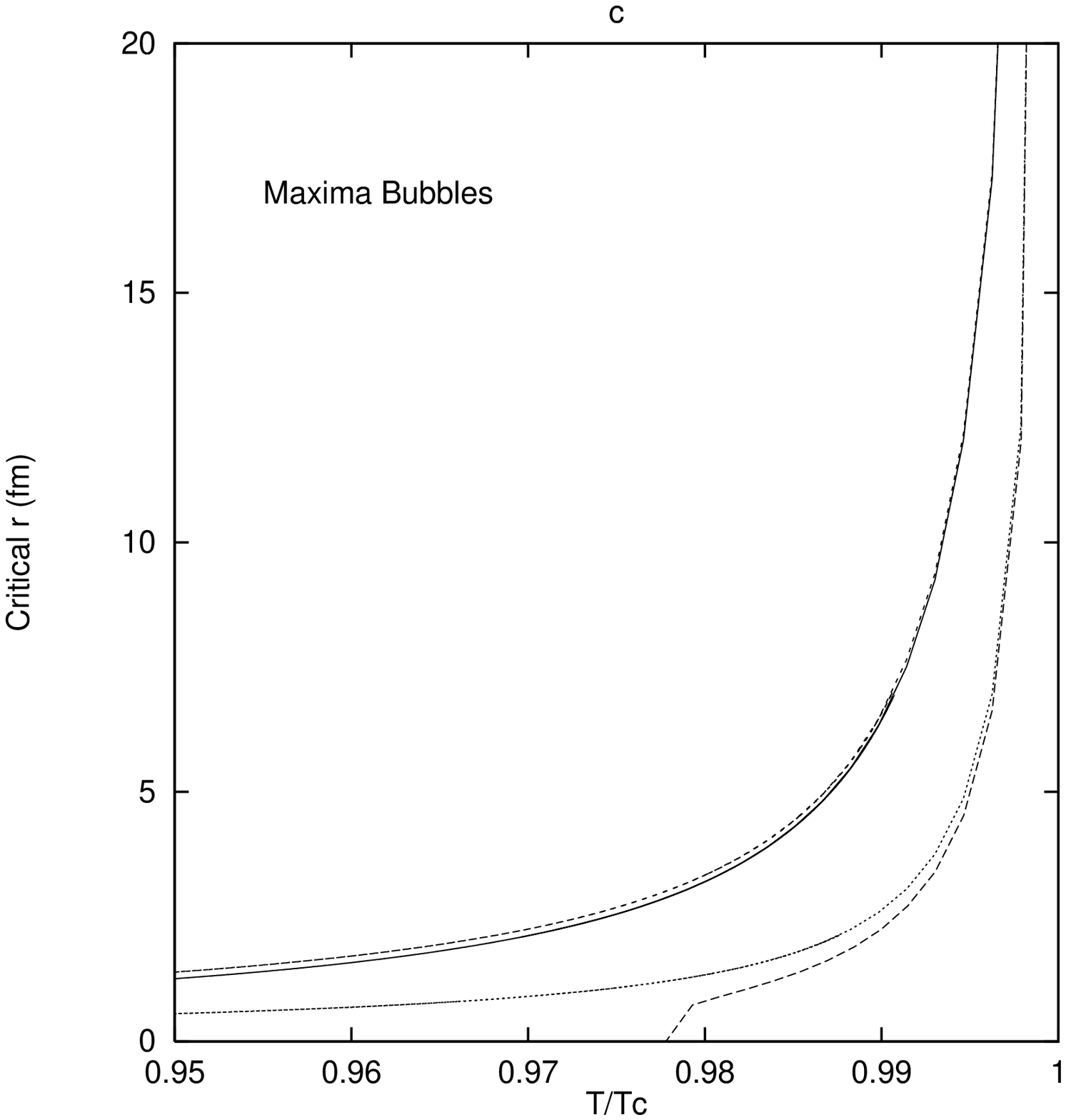}
\hskip 1cm
\epsfxsize=5cm\epsfysize=6cm\epsfbox{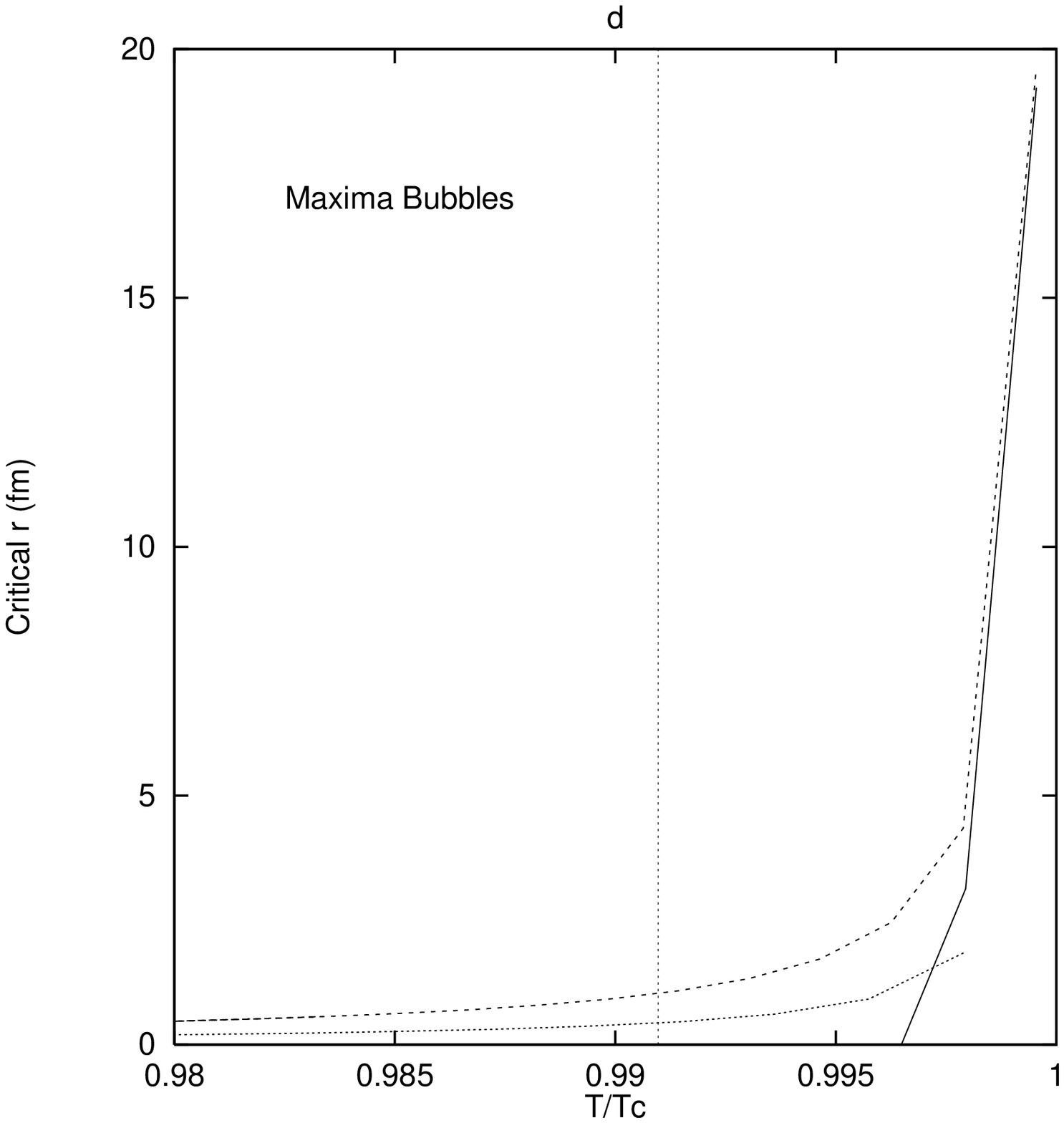}}
\caption{Upper panel shows critical bubble radius $r_c$ in fm as a function of $T/T_c$. They are the
minimum free energy equilibrium critcal bubbles. The solid and long dashed curves are both 
with $\gamma $ correction. These bubbles do not exist without the curvature 
correction. Figures 4a and 4b are labelled as in fig.1a and 1b respectively.
Lower panel shows critical bubble radius $r_c$ in fm as a function of $T/T_c$. They are the
maximum free energy expanding critcal bubbles. Figures 4c and 4d are labelled as in 
fig.1a and 1b respectively.}
\end{figure}
\pagebreak
\begin{figure}[ht]
\vspace*{1cm}
\centerline{
\epsfxsize=5cm\epsfysize=6cm\epsfbox{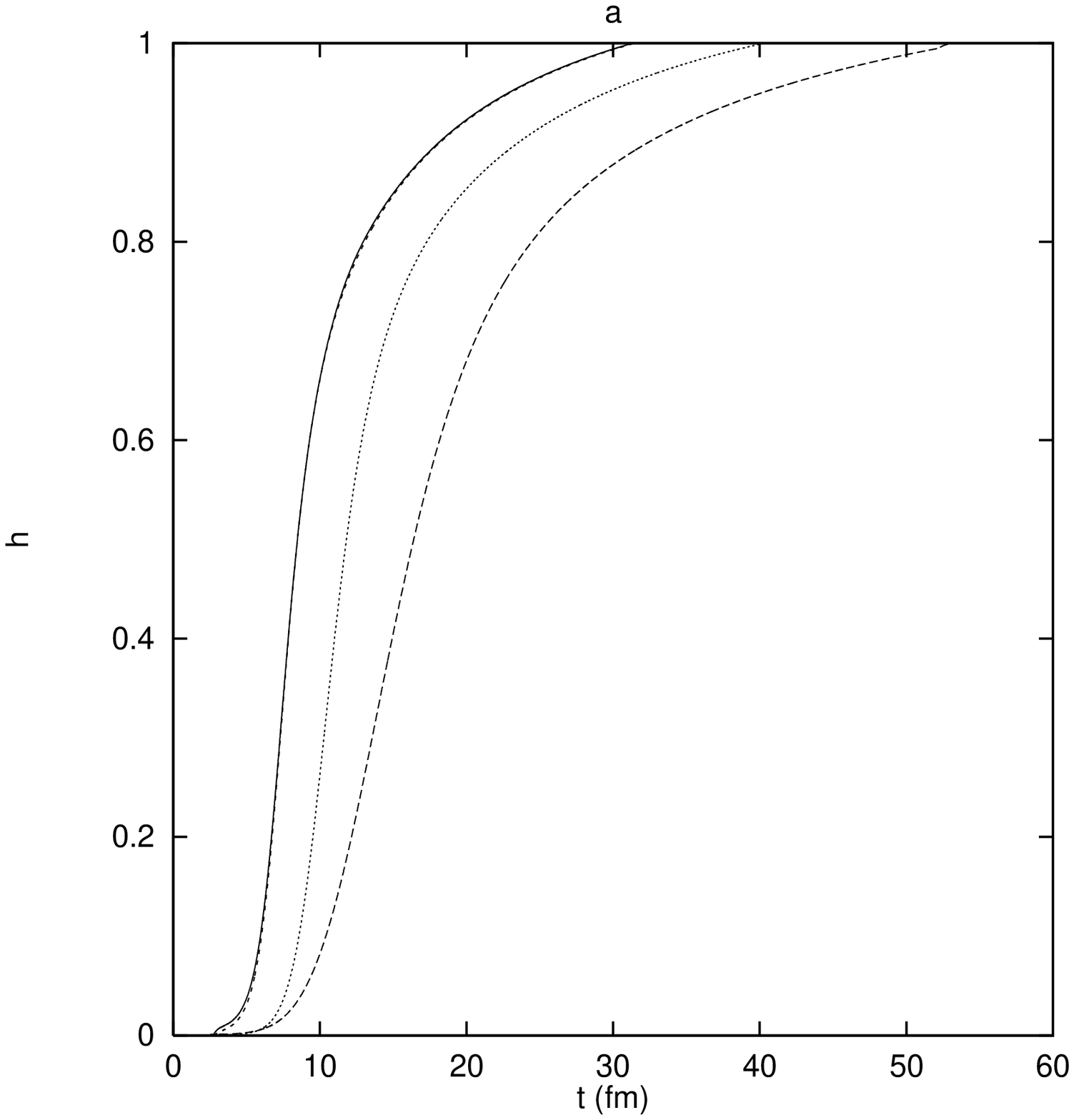}
\hskip 1cm
\epsfxsize=5cm\epsfysize=6cm\epsfbox{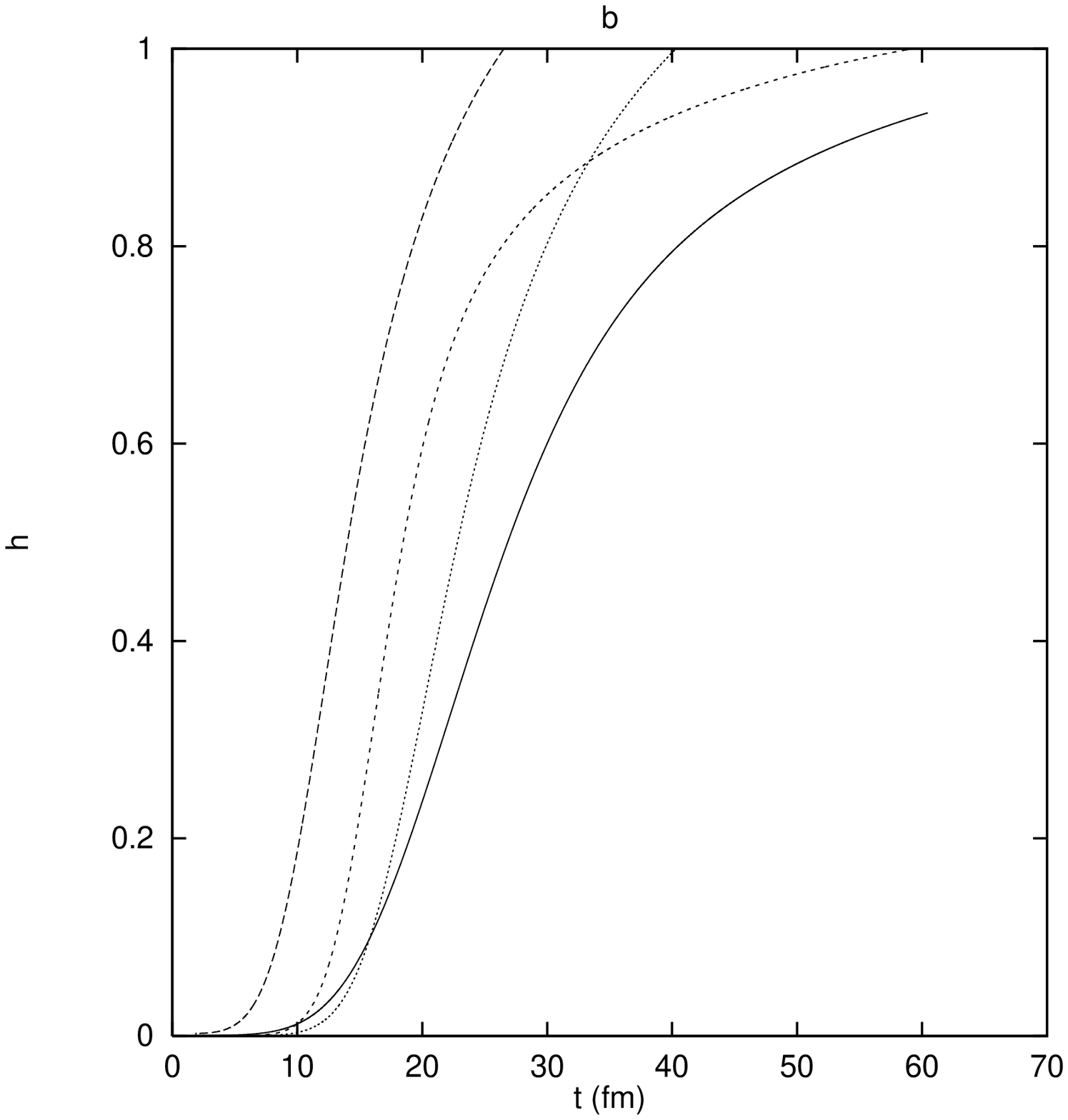}}
\caption{The hadron fraction $h$ as a function of time $t$ in fm. 
Curves in fig.5a and 5b 
are labelled as in fig. 1a and 1b respectively.}
\end{figure}
\end{document}